\documentclass{rspublic}
\begin{document}

\newcommand{\half}{\mbox{$\textstyle \frac{1}{2}$}}
\newcommand{\roothalf}{\mbox{$\textstyle \frac{1}{\sqrt{2}}$}}
\newcommand{\rootthird}{\mbox{$\textstyle \frac{1}{\sqrt{3}}$}}
\newcommand{\rootsixth}{\mbox{$\textstyle \frac{1}{\sqrt{6}}$}}
\newcommand{\threetwo}{\mbox{$\textstyle \frac{3}{2}$}}
\newcommand{\twothirds}{\mbox{$\textstyle \frac{2}{3}$}}
\newcommand{\third}{\mbox{$\textstyle \frac{1}{3}$}}
\newcommand{\quat}{\mbox{$\textstyle \frac{1}{4}$}}
\newcommand{\octa}{\mbox{$\textstyle \frac{1}{8}$}}

\title[Unusual quantum states:]{Unusual quantum states:
nonlocality, entropy, Maxwell's daemon, and fractals}

\author[Bender, Brody, $\&$
Meister]{Carl~M.~Bender\footnote{Permanent address: Department of
Physics, Washington University, St. Louis MO 63130, USA.},
Dorje~C.~Brody and Bernhard~K.~Meister}

\affiliation{Blackett Laboratory, Imperial College, London SW7
2BZ, UK }

\date{\today}
\maketitle
\input{psfig.sty}

\begin{abstract}{energy conservation; nonlocality; quantum
measurement; Bell inequality; fractals; von Neumann entropy;
Maxwell's daemon} This paper analyses the mathematical properties
of some unusual quantum states that are constructed by inserting
an impenetrable barrier into a chamber confining a single
particle. If the barrier is inserted at a fixed node of the wave
function, then the energy of the system is conserved. After
barrier insertion, a measurement is made on one side of the
chamber to determine if the particle is physically present. The
measurement causes the wave function to collapse, and the energy
that was contained in the subchamber in which the particle is now
absent transfers instantaneously to the other subchamber in which
the particle now exists. This thought experiment constitutes an
elementary example of an EPR experiment based on energy
conservation rather than momentum or angular momentum
conservation. A more interesting situation arises when one inserts
the barrier at a point that is not a fixed node of the wave
function because this process changes the energy of the system;
the faster the barrier is inserted, the greater the change in the
energy. At the point of a sudden insertion the energy density
becomes infinite; this energy instantly propagates across the
subchamber and causes the wave function to become fractal. If an
energy measurement is carried out on such a fractal wave function,
the resulting mixed state has finite nonzero entropy. Fractal
mixed states having unbounded entropy are also constructed and
their properties are discussed. For an adiabatic insertion of the
barrier, Landauer's principle is shown to be insufficient to
resolve the apparent violation of the second law of thermodynamics
that arises when a Maxwell daemon is present. This problem is
resolved by calculating the energy required to insert the barrier.
\end{abstract}

\section{Introduction}
\label{s1}

The Hamiltonian describing a particle of mass $m$ trapped in a
one-dimensional infinite square well of width $L$ is
\begin{eqnarray}
H=-\frac{\hbar^2}{2m}\frac{\rd^2}{\rd x^2} \label{eq:1.1}
\end{eqnarray}
with the boundary conditions that the wave function $\psi(x)$
vanish at $x=0$ and $x=L$. The stationary states and the
corresponding energy eigenvalues are
\begin{eqnarray}
\phi_n(x)=\sqrt{\frac{2}{L}}\sin\left(\frac{n\pi x}{L}\right)
\quad {\rm and}\quad E_n=\frac{\pi^2\hbar^2n^2}{2mL^2}.
\label{eq:1.2}
\end{eqnarray}

The main purpose of this paper is to investigate the effect of
inserting an impenetrable barrier at a point $x_0=L-\delta$ inside
the potential well. A related investigation for $x_0=L/2$ using
high-temperature quantum states was done by Zurek (1986) in his
study of Maxwell's daemon. Our objective here is to provide a
thorough analysis of the effects of barrier insertion for
arbitrary $x_0$. Inserting an impenetrable barrier divides the
chamber into two noninteracting subchambers. If there is a
particle in the original undivided chamber, then after inserting
the barrier, this particle can only be present in a classical
sense in one of the two subchambers. Quantum mechanically, after
the barrier is inserted, the wave function of the particle remains
nonzero in {\it both} subchambers. We thus consider what happens
if we perform a measurement that determines the presence or
absence of the particle in a given subchamber. Such a measurement
requires a probe, which we treat classically.\footnote{An example
of a quantum probe would be a Heisenberg microscope in which a
dense beam of photons is aimed at the subchamber. The presence of
the particle would then be confirmed by the scattering of photons
out of the beam. A probe of this kind for a particle trapped in a
harmonic well was envisaged by Dicke (1981), who considered the
possibility of quantum effects associated with the photons in the
beam. The problem with a photon beam probe is that in order for
the particle in the box to interact with the photons, it must be
electrically charged. This allows the particle to absorb energy
from the photon beam and to radiate energy into the photon beam.
Although the consequences of such a quantum probe are interesting,
we do not address this additional complexity here.}

{\it The probe:} As a simple classical probe we use movable walls
at the ends of each of the subchambers (these walls are treated as
classical objects). A quantum particle in a subchamber exerts a
force on the walls of the chamber (Bender {\it et al}. 2002). By
pushing on a wall we either detect a force or not and thus we
measure the presence or the absence of the particle.

For a given initial state $\psi(x,0)$ of the system, there are two
possibilities for the insertion point $x_0$ of the barrier;
namely, the fixed nodes of the wave function and all other
locations. By a {\it fixed node} we mean a point $x_0$ for which
the time-dependent wave function satisfies $\psi(x_0,t)=0$ for all
$t\geq0$.

{\it Case 1: Inserting an impenetrable barrier at a fixed node.}
Because the wave function vanishes at the insertion point, the
system cannot detect and therefore does not resist the insertion
of a barrier. Thus, inserting a barrier does not change the energy
of the system. After the insertion, the initial energy is
distributed between the two subchambers, one to the left and the
other to the right of the barrier. We remark that the wave
function cannot collapse to zero in either of the subchambers
during the barrier insertion because, if it did, the entropy of
the system would be reduced without doing any work.

{\it Case 2: Inserting an impenetrable barrier at a point that is
not a fixed node.} In this case inserting a barrier changes the
energy by an amount that depends on the rate at which the barrier
is inserted. The time evolution of the system after the insertion
of the barrier depends on how rapidly the barrier is inserted.

In \S\ref{s2} and \ref{s3} we examine Case 1 and study the effect
of a measurement that determines which side of the inserted
barrier the particle is present. Based on the conventional
collapse hypothesis of quantum mechanics, we argue that the energy
in the subchamber where the particle is found to be absent
transfers instantaneously to the other subchamber even if the two
subchambers are first separated adiabatically by a great distance.
This phenomenon of energy transfer is analogous to the
instantaneous transfer of angular momentum in an EPR experiment.

When the barrier is not inserted at a fixed node (Case 2), the
analysis becomes more elaborate. We consider in \S\ref{s4} what
happens when the insertion is carried out instantaneously. After
the insertion it is natural to expand the initial wave function
$\psi(x,0)$ in the regions $[0,L-\delta)$ and $(L-\delta,L]$
separately as Fourier series. The orthonormal basis elements in
each of the two regions are eigenstates of the Hamiltonians for
each of the subchambers. It is shown that a naive calculation of
the energy in terms of the expansion coefficients of the wave
function gives a divergent expression. This happens because while
the series expansion of the initial wave function is convergent,
it is not uniformly convergent and thus the series exhibits the
Gibbs phenomenon.

Calculating the expectation value of the Hamiltonian requires that
the series representing the wave function be twice differentiated.
However, one cannot differentiate a nonuniformly convergent
Fourier series term by term. By introducing a simple
renormalisation scheme we overcome the nonuniform convergence and
obtain a finite answer for the expectation value of the
Hamiltonian on each side of the barrier. When the resulting
expectation values are added, we find that this agrees with the
initial energy of the system. However, this calculation still does
not determine the total energy because the result does not include
the energy density at the insertion point $x_0$. At $x_0$ the wave
function is discontinuous, and consequently the energy density at
$x_0$ is infinite at the time $t=0^+$ immediately after the
barrier is inserted. Evidently, it requires an infinite amount of
work to insert an impenetrable barrier at a point where there is
no fixed node. The effect of a measurement to determine which
subchamber contains the particle after a rapid insertion of the
barrier is considered in \S\ref{s5}.

When an initial state created by the instantaneous insertion of a
barrier evolves in time, the discontinuity at $x_0$ instantly
disappears, and the wave function becomes continuous and
square-integrable. However, an infinite amount of energy is stored
inside the well and, as a result, the wave function becomes
fractal. Thus, the wave function, expressed as a linear
combination of the energy eigenstates, no longer belongs to the
domain of the Hamiltonian although it remains in the domain of the
unitary time evolution operator. In \S\ref{s6} we show that the
quantum states encountered here resemble the fractal wave
functions introduced by Berry (1996).

In \S\ref{s7} we consider an adiabatic insertion of the barrier. A
slow insertion requires only a finite amount of energy, and in the
adiabatic limit this energy can be calculated explicitly. In
\S\ref{s8} we contrast our results on adiabatic insertion with the
work of Zurek (1986) on Maxwell's daemon. We show that the
argument of Zurek about memory erasure of the daemon to support
the second law of thermodynamics is insufficient when the barrier
is not inserted at the centre of the well.

If we measure the energy of the particle described by a fractal
wave function created by an instantaneous insertion of the
barrier, the resulting mixed state has finite nonzero von Neumann
entropy. We extend this analysis in \S\ref{s9} by constructing a
new class of fractal wave functions for which the density matrix
of the state has unbounded entropy. Such a state might be obtained
by an instantaneous compression, as opposed to an instantaneous
division, of the chamber.

In gravitational physics the black-hole surface area provides a
bound on entropy. Thus, the existence of a quantum state having
unbounded von Neumann entropy in a finite-size quantum well seems
to imply that nonrelativistic quantum mechanics can violate the
entropy bound of gravitational physics (which in itself would not
be surprising). However, the infinite-entropy quantum state
introduced in \S\ref{s9} also possesses infinite energy. More
generally, we argue in \S\ref{s10} that for a broad class of
potentials, any quantum state having infinite entropy must also
have infinite energy and/or infinite characteristic size. In
\S\ref{s11} we discuss the consequences of separating the two
subchambers that are formed after a barrier is inserted. We
conclude in \S\ref{s12} with some open issues.

\section{Barrier insertion at a fixed node}
\label{s2}

\begin{figure}
{\centerline{\psfig{file=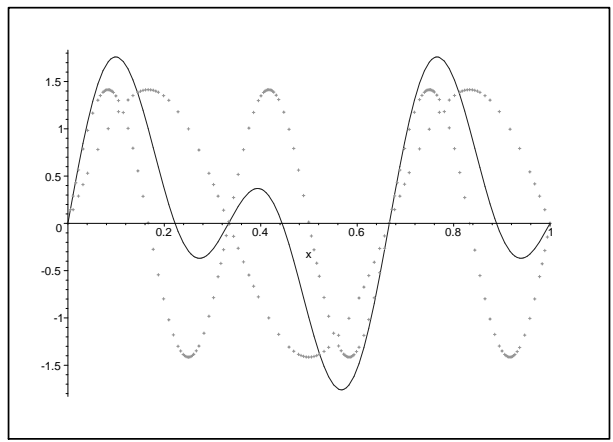,width=8cm,angle=270}}}
\caption{\label{F1} The plot of the initial wave function in
(\ref{eq:2.1}),
$\psi(x,0)=\frac{1}{\sqrt{2}}[\phi_3(x)+\phi_6(x)]$, for $L=1$.
The dotted lines represent the two eigenstates $\phi_3(x)$ and
$\phi_6(x)$. The nodes that are common to both eigenstates are the
fixed nodes of $\psi(x,t)$. The other nodes of $\psi(x,0)$ are
unstable in the sense that $\psi(x,t)$ does not continue to vanish
at these points.}
\end{figure}

In this section we determine the effect of inserting an
impenetrable barrier at a fixed node of the initial wave function.
Any of the energy eigenstates other than the ground state has
fixed nodes. Furthermore, a superposition of energy eigenstates
can also have fixed nodes. As an example, let us choose the
initial state
\begin{eqnarray}
\psi(x,0)=\roothalf[\phi_3(x)+ \phi_6(x)]. \label{eq:2.1}
\end{eqnarray}
Using (\ref{eq:1.2}) we see that the initial energy of the system
described by (\ref{eq:2.1}) is $\frac{45}{2}(\pi^2\hbar^2/2mL^2)$.
As shown in Fig.~\ref{F1}, this initial state has nodes at
$x/L=0,\frac{2}{9}$, $\frac{1}{3}$, $\frac{4}{9}$, $\frac{2}{3}$,
$\frac{8}{9}$, $1$. However, the nodes at $\frac{2}{9}$,
$\frac{4}{9}$, $\frac{8}{9}$ are not fixed because the value of
the time-dependent wave function $\psi(x,t)$ does not remain zero
at these points.

Let us insert an impenetrable barrier at one of the fixed nodes,
say $x_0=\frac{2}{3}L$, at time $t=0$. The value of $L$ is
irrelevant to the discussion here, so we set $L=1$ in this
section. Because the inserted barrier is impenetrable, the two
subregions can be treated separately. In fact, these two regions
can even be moved apart without affecting the states if the
separation is done adiabatically. Nevertheless, the state of the
system as a whole is described by a single entangled wave
function.

The Hamiltonian describing each subregion is the same as that in
(\ref{eq:1.1}) with the new boundary conditions being that the
wave functions must vanish at $x=0$ and at $x=2/3$ in the first
subregion and at $x=2/3$ and at $x=1$ in the second subregion. The
eigenstates
\begin{eqnarray}
\chi_n(x)=\sqrt{3}\sin\left(\threetwo\pi nx\right) \quad {\rm and}
\quad \eta_n(x)=\sqrt{6}\sin\left[3\pi n(x-\twothirds)\right]
\label{eq:2.2}
\end{eqnarray}
of the Hamiltonians in these two regions constitute orthonormal
sets of basis elements in each of the regions $[0,\frac{2}{3}]$
and $[\frac{2}{3},1]$. Therefore, the initial state is
\begin{eqnarray}
\psi(x,0) = \left\{ \begin{array}{ll} \rootthird \left[ \chi_2(x)
+ \chi_4(x) \right] & (0\leq x\leq \frac{2}{3}), \\ \rootsixth
\left[ \eta_1(x) + \eta_2(x) \right] & (\frac{2}{3}\leq x\leq1).
\end{array} \right. \label{eq:2.3}
\end{eqnarray}
Note that the wave function is not normalised in either region.

Let us calculate the expectation value of the Hamiltonian. The
energy eigenvalues in each region are $E_n^\chi= \frac{9}{4}
n^2(\pi^2\hbar^2/2m)$ and $E_n^\eta=9n^2 (\pi^2\hbar^2/2m)$,
respectively. Using the basis decomposition (\ref{eq:2.3}) the
energy expectation values for each side of the barrier are
$E_\chi=\frac{45}{3}(\pi^2\hbar^2/2m)$ and $E_\eta=\frac{45}{6}
(\pi^2\hbar^2/2m)$. Thus,
$E_\chi+E_\eta=\frac{45}{2}(\pi^2\hbar^2/2m)$, which is just the
initial energy of the system. Because the energy expectation is a
constant under unitary time evolution the expectation value
obtained from the time-dependent wave function $\psi(x,t)$ gives
the identical result. Analogous results hold for all other initial
pure states, and we conclude that inserting an impenetrable
barrier at a fixed node requires no energy.

\section{Wave-function collapse and nonlocal energy transfer}
\label{s3}

What happens if we observe whether the particle is present in one
of the two subregions, say $[\frac{2}{3},1]$ after we have
inserted an impenetrable barrier? As discussed in \S\ref{s1}, the
thought experiment that we have in mind consists of detecting
whether there is a force on the wall of the subchamber. If the
particle is trapped in the range $[\frac{2}{3},1]$, then we can
detect the force that the particle exerts on the wall of the
subchamber.

Our concern here is how such a measurement affects the energy of
the system. We have established in the previous section that prior
to the measurement the energy is distributed among the two
separate subregions, even though there is only one particle. Let
us suppose that the measurement confirms the absence of the
particle in the range $[\frac{2}{3},1]$. Such an outcome occurs
with probability $\frac{1}{3}$. After this measurement we can {\it
infer} that the value of $E_\eta$ is now zero. This appears to be
paradoxical unless we can explain where the energy has
gone.\footnote{The situation considered here is similar to a
double-slit experiment in which a particle is not localised at one
of the two slits; localising the particle at one slit destroys the
entanglement, which is observed as an interference pattern.}

How does this measurement affect the normalisation of the wave
function? Recall that the wave function $\rootthird[
\chi_2(x)+\chi_4(x)]$ of (\ref{eq:2.3}), which has support on
$[0,\frac{2}{3}]$, is not normalised to unity because prior to the
measurement this wave function did not represent the state of the
entire system. However, once the absence of the particle is
confirmed by the measurement, the state of the system collapses to
a new state having support only on $[0,\frac{2}{3}]$. If we follow
the standard approach to quantum measurement theory (cf. L\"uders
1951), then the resulting state is given simply by $\rootthird
[\chi_2(x)+\chi_4(x)]$, divided by its norm (the minimum
projection), so that the wave function, having support on
$[0,\frac{2}{3}]$, is now normalised.

The squared norm of the state $\rootthird[\chi_2(x)+\chi_4(x)]$ is
$\frac{2}{3}$, so after the absence of the particle in the region
$[\frac{2}{3},1]$ has been confirmed, the expectation value of the
Hamiltonian in the region $[0,\frac{2}{3}]$ becomes
$\frac{45}{2}(\pi^2\hbar^2/2m)$, which is the initial energy of
the system. The argument outlined here generalises to the case of
an arbitrary initial state for which a barrier is inserted at any
fixed node.

To summarise, we insert an impenetrable barrier in the potential
well where there is a fixed node, and we calculate the expectation
values of the Hamiltonian on each side of the barrier. The sum of
these expectation values equals the initial energy of the system.
Then, we perform a measurement and observe the presence or absence
of the particle on one side of the barrier. If the particle is
absent, then the energy $E_\eta$ in the subchamber is transferred
to the other chamber instantaneously so that the total energy
remains conserved.

A similar situation was addressed by Dicke (1981), who considered
the use of a Heisenberg microscope probe for a particle trapped in
a harmonic potential. In his thought experiment, the absence of
the particle in a given region is confirmed by the lack of
scattered photons. However, a conceptual difficulty arises because
the energy of the particle is affected even though the particle
has not interacted with the photon beam. As a possible resolution,
Dicke offered the explanation that the quantum state of the
unobserved particle (the particle that is absent) is altered by
the absorption and emission of photons. What seems difficult to
explain, however, is the mechanism by which the energy is
transferred.

The difficulty in understanding the phenomenon of energy transfer
is, in fact, more severe than Dicke envisaged. In the experiment
considered by Dicke, there is no physical obstacle between the
photon beam used as a probe and the region where the particle is
trapped, whereas in our thought experiment there is an
impenetrable potential barrier separating these two regions. We
can even separate the two chambers adiabatically as far apart as
we wish before performing the position measurement. When the
result of the position measurement is obtained, there is an
instantaneous transfer of energy from one chamber to a remotely
separated chamber.

This thought experiment is an elementary example of an EPR
experiment (Einstein {\it et al}. 1935) except that here there is
only one particle. The EPR analysis uses momentum conservation to
reveal the peculiar feature of nonlocality in quantum mechanics. A
more familiar version of the EPR experiment based on a singlet
state of a pair of spin-$\frac{1}{2}$ particles relies on
conservation of angular momentum to reveal nonlocality (Bohm
1951). In contrast, in this paper we implicitly accept the
nonlocality of quantum mechanics and demonstrate the conservation
of energy. We could just as well have assumed energy conservation
and have used it to demonstrate nonlocality.

\section{Insertion of the barrier at a nonnodal point}
\label{s4}

\begin{figure}
{\centerline{\psfig{file=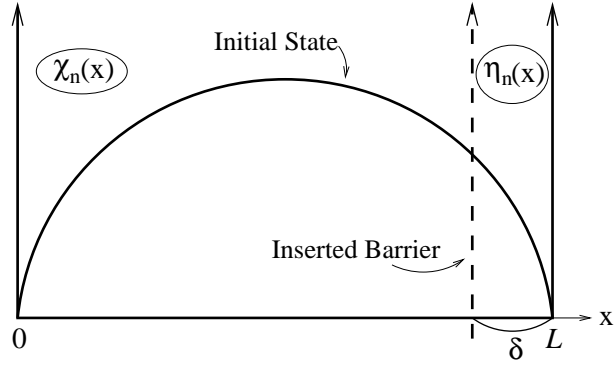,width=8cm,angle=270}}}
\caption{\label{F2} Insertion of an impenetrable potential barrier
in the square well, where the initial state of the system is the
ground state. The initial state can be expanded separately in the
two subregions using the basis functions $\chi_n(x)$ and
$\eta_n(x)$.}
\end{figure}

Let us now consider the instantaneous insertion of a barrier where
there is no fixed node. For simplicity, we take the initial wave
function to be the ground state $\psi(x,0)=\phi_1(x)$ for
$x\in[0,L]$, and we insert an impenetrable wall at $x=L-\delta$.
This setup is illustrated in Fig.~\ref{F2}. The probability of
trapping the particle in the outer chamber (between $L-\delta$ and
$L$) is determined by the integral
$p=\int_{L-\delta}^L\psi^2(x,0)\rd x$, and for small values of
$\epsilon=\delta/L$ we have the expansion
\begin{eqnarray}
p(\epsilon)\sim\frac{2}{3}\pi^2\epsilon^3 - \frac{2}{15}\pi^4
\epsilon^5 + {\cal O}(\epsilon^7). \label{eq:5.1}
\end{eqnarray}
If the particle in the box behaved classically, then the
probability distribution for the location of the particle would be
uniform because the velocity of the particle would be constant.
Quantum mechanically, for a highly excited state $\phi_n(x)$
(large $n$), the particle behaves classically. Indeed, for large
$n$ we have $p(\epsilon)\sim \epsilon$. The cubic behaviour in
(\ref{eq:5.1}) is a characteristic feature of low-energy quantum
states.

Because we have placed an infinite potential barrier at the point
$x=L-\delta$, the two subchambers can now be treated separately,
as in the previous example. Thus, we consider an orthonormal set
of basis elements on each side of the wall:
\begin{eqnarray}
\chi_n(x) = \sqrt{\frac{2}{L-\delta}} \sin\left( \frac{n\pi x}
{L-\delta} \right) \quad (x\in[0,L-\delta]) \label{eq:5.2}
\end{eqnarray}
and
\begin{eqnarray}
\eta_n(x) = \sqrt{\frac{2}{\delta}} \sin\left( \frac{n\pi
(x-L+\delta)} {\delta} \right) \quad (x\in[L-\delta,L]).
\label{eq:5.3}
\end{eqnarray}
Using these bases we can expand the initial wave function with
coefficients given by $a_n=\int_0^{L-\delta}\psi(x) \chi_n(x)\rd
x$ and $b_n=\int_{L-\delta}^L\psi(x) \eta_n(x)\rd x$, in each of
the two regions. Straightforward calculations show that
\begin{eqnarray}
a_n = \frac{2n(-1)^{n+1}\sqrt{1-\epsilon}\sin(\pi\epsilon)}
{\pi[n+(1-\epsilon)][n-(1-\epsilon)]} \quad {\rm and} \quad b_n =
\frac{2n\sqrt{\epsilon}\sin(\pi\epsilon)}
{\pi(n+\epsilon)(n-\epsilon)}, \label{eq:5.4}
\end{eqnarray}
where $\epsilon=\delta/L$. Therefore, the initial wave function
for the region $x\in[0,L-\delta)$ is
\begin{eqnarray}
\psi_{\chi}(x,0) = \sum_{n=1}^\infty a_n \chi_n(x), \label{eq:5.5}
\end{eqnarray}
where $\psi_\chi(x,0)=\psi(x,0)$ for $x\in[0,L-\delta)$ and
$\psi_\chi(x,0)=0$ otherwise.

The notation $[0,L-\delta)$ indicates that the Fourier expansion
of the initial wave function converges {\it pointwise} in the
half-open interval that contains the left endpoint $0$ but not the
right endpoint $L-\delta$. In general, a Fourier sine series
converges pointwise and uniformly to a continuously differentiable
function satisfying homogeneous boundary conditions at the two
endpoints. If a continuously differentiable function vanishes at
one endpoint (say, the left endpoint) but not at the other (say,
the right endpoint), then the Fourier sine series converges
pointwise to the function everywhere except at the right endpoint,
where we observe the rapid oscillation known as the {\it Gibbs
phenomenon} and nonuniform convergence of the Fourier series.

In the region $x\in(L-\delta,L]$ we have an analogous expression
for $\psi_\eta(x)$ in terms of $b_n$ and $\eta_n(x)$. Because of
the symmetry associated with the problem, the results for
$x\in(L-\delta,L]$ are identical to those for $x\in[0,L-\delta)$,
under the substitution $\epsilon\to 1-\epsilon$. Therefore, we
need only analyse the range $[0,L-\delta)$ from which we infer the
corresponding results in $(L-\delta,L]$.

The expectation value $E_\chi$ of the Hamiltonian $H$ in
(\ref{eq:1.1}) for $x\in[0,L-\delta)$ is given by
\begin{eqnarray}
E_\chi = \int_0^{L-\delta}\psi_{\chi}^*(x,0)H\psi_{\chi}(x,0)\rd
x. \label{eq:5.6}
\end{eqnarray}
Note that the wave function $\psi_\chi(x)$ is not normalised to
unity because it does not represent the totality of the system.

If we substitute (\ref{eq:5.5}) in (\ref{eq:5.6}) to calculate the
energy of the subsystem in the range $[0,L-\delta)$, then we
obtain the divergent series representation for $E_\chi$:
\begin{eqnarray}
\sum_{n=1}^\infty a_n^2 \frac{\pi^2\hbar^2n^2} {2m(L-\delta)^2}.
\label{eq:5.7}
\end{eqnarray}
This series diverges because $a_n^2\sim1/n^2$ for large $n$.
However, this divergence does not imply that the energy of the
particle in the region $[0,L-\delta)$ is infinite. This apparent
paradox arises because the series (\ref{eq:5.5}), while it
converges to $\psi_\chi(x,0)$, is not uniformly convergent.
Consequently, when the differential operator $H$ acts on
$\psi_\chi(x,0)$ in (\ref{eq:5.6}), we cannot differentiate the
series term by term to obtain (\ref{eq:5.7}). The correct way to
calculate $E_\chi$ relies on expressing the initial wave function
$\psi_\chi(x,0)$ of (\ref{eq:5.5}) in the form
\begin{eqnarray}
\psi_{\chi}(x,0) &=& \sum_{n=1}^\infty a_n \chi_n(x) \nonumber \\
&=& \alpha \sum_{n=1}^\infty (-1)^{n+1} \sin\left(\frac{n\pi
x}{L-\delta}\right) \left[ \frac{n}{n^2-\beta^2}-\frac{1}{n}+
\frac{1}{n}\right] \nonumber \\ &=& \alpha \beta^2
\sum_{n=1}^\infty \frac{(-1)^{n+1}}{n(n^2-\beta^2)}\sin
\left(\frac{n\pi x}{L-\delta}\right) + \alpha \sum_{n=1}^\infty
\frac{(-1)^{n+1}}{n}\sin\left(\frac{n\pi x}{L-\delta}\right)
\nonumber \\ &=& \alpha \beta^2 \sum_{n=1}^\infty
\frac{(-1)^{n+1}}{n(n^2-\beta^2)}\sin \left(\frac{n\pi
x}{L-\delta}\right) + \frac{\alpha\pi}{2(L-\delta)}x ,
\label{eq:5.9}
\end{eqnarray}
where $\alpha=(2\sqrt{2}/\pi\sqrt{L})\sin(\pi\epsilon)$ and
$\beta=1-\epsilon$. We recognise that the second term in the final
step in (\ref{eq:5.9}) is the Fourier series representation of the
function $x$. The Gibbs phenomenon arises here because the initial
wave function does not satisfy the homogeneous boundary conditions
obeyed by each of the Fourier eigenmodes. After subtracting the
function proportional to $x$, the remaining function satisfies
homogeneous boundary conditions and the Gibbs phenomenon
evaporates.

Let us explain in detail the idea behind the formal manipulation
in (\ref{eq:5.9}). First, we note that the large-$n$ asymptotic
behaviour of the Fourier coefficients is $a_n\sim1/n$, which gives
rise to the divergent series (\ref{eq:5.7}). Therefore, in
(\ref{eq:5.9}) we isolate the contribution of order $1/n$ in the
summand so that the reminder has a $1/n^3$ behaviour for large $n$
and the series converges rapidly. The leading $1/n$ behaviour
comes from the Fourier coefficients of the function $x$. Although
the series expansion of $x$ is not uniformly convergent, we can
explicitly perform the summation before applying the differential
operator. In effect, this renormalisation procedure eliminates the
Gibbs phenomenon, and we obtain a finite answer. Indeed, if we
substitute the final line of (\ref{eq:5.9}) into (\ref{eq:5.6}),
after some algebra we find that the renormalised energy is
\begin{eqnarray}
E_\chi = \frac{\pi^2\hbar^2}{2mL^2} \frac{(1-\epsilon)
\sin^2(\pi\epsilon)}{\pi^2}
\left(\Psi_1(\epsilon)+\Psi_1(2-\epsilon)+
\frac{\Psi_0(2-\epsilon)-\Psi_0(\epsilon)}{1-\epsilon}\right).
\label{eq:5.10}
\end{eqnarray}
Here, $\Psi_0(u)$ denotes the digamma function
$\Psi_0(u)=\rd\ln\Gamma(u)/\rd u$ and $\Psi_1(u)=\rd\Psi(u)/\rd u$
is the first derivative of the digamma (first polygamma) function.
The result in (\ref{eq:5.10}) is exact and holds for all values of
$\epsilon\in[0,1]$.

The expectation value $E_\eta$ of the Hamiltonian in the range
$(L-\delta,L]$ is obtained by substituting $1-\epsilon$ for
$\epsilon$ in (\ref{eq:5.10}). Therefore, the expectation value of
the Hamiltonian in the two half-open regions $[0,L-\delta)$ and
$(L-\delta,L]$ is $E_\chi+E_\eta$. From the properties of the
digamma and polygamma functions we find that
\begin{eqnarray}
E_\chi + E_\eta = \frac{\pi^2\hbar^2}{2mL^2} \label{eq:5.11}
\end{eqnarray}
for all values of $\epsilon$. As this is just the initial energy
of the system, one might conclude that energy is conserved during
the insertion of the wall. However, the expression (\ref{eq:5.11})
does not take into account the energy density at the point
$x=L-\delta$. Because the barrier is assumed impenetrable, the
wave function must vanish at $L-\delta$. Therefore, at time $t=0$
the wave function is discontinuous at $L-\delta$ and the energy
density is infinite. Thus, an instantaneous insertion of the
barrier at a nonnodal location requires an infinite amount of
energy.

\section{Effect of a position measurement}
\label{s5}

Let us study, as we did in \S\ref{s3}, the effect of a measurement
that determines the presence or absence of the particle after the
instantaneous insertion of the barrier at $t=0$. We are concerned
with the energy initially contained in the chamber and thus we
also perform the subsequent position measurement at time $t=0$.

Without the measurement the conservation law (\ref{eq:5.11}) holds
in the open region $[0,L]-\{L-\delta\}$. However, if the
measurement is performed and confirms the absence of the particle
in the range $(L-\delta,L]$, then after this measurement we can
infer that the value of $E_\eta$ is now zero. Recall that the wave
function $\psi_\chi(x,0)$, having support on $[0,L-\delta)$, is
not normalised to unity. However, once the absence of the particle
is confirmed by measurement, then the state of the system will
collapse into a new state having support on $[0,L-\delta)$, and we
must redetermine the normalisation of the wave function as we did
in \S\ref{s3}.

\begin{figure}
{\centerline{\psfig{file=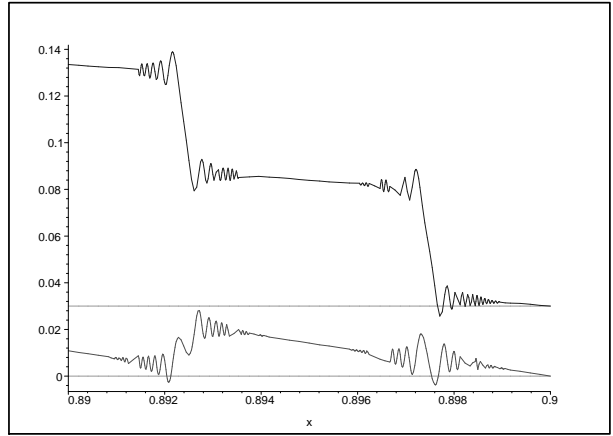,width=8cm,angle=270}}}
\caption{\label{F3} Real (top curve) and imaginary (bottom curve)
part of the wave function for a particle in the box, approximated
by its Fourier expansion for the first 25,000 terms. (The real
part has been shifted upwards by 0.03 to distinguish it from the
imaginary part.) We set $L=1$\AA, $\delta=0.1$\AA, and chosen the
mass $m_{\rm e}$ of the particle to be the mass of the electron.
The plot shows the wave function in the interval
0.89\AA~$\sim$~0.9\AA, at time $t=10^{-17}$ seconds. Evidently,
the values of the real and imaginary parts of the wave function
rapidly approach zero at $x_0=0.9$\AA, where the barrier bas been
inserted.}
\end{figure}

The squared norm of $\psi_\chi(x,0)$ can be calculated in two
ways, either by integrating the square of $\psi_\chi(x,0)$ or by
summing the squares of the expansion coefficients $a_n$. The
result of the former calculation is
\begin{eqnarray}
\int_0^{L-\delta} \psi_{\chi}^2(x,0) \rd x = \frac{1}{\pi}
\sin(\pi\epsilon)\cos(\pi\epsilon) + (1-\epsilon), \label{eq:6.1}
\end{eqnarray}
whereas the latter calculation gives
\begin{eqnarray}
\sum_{n=1}^\infty a_n^2 &=& \frac{4}{\pi^2}(1-\epsilon)\sin^2(\pi
\epsilon) \sum_{n=1}^\infty \frac{n^2}{[n^2-(1-\epsilon)^2]^2}
\nonumber \\ &=& \frac{(1-\epsilon) \sin^2(\pi\epsilon)}{\pi^2}
\left( \Psi_1(\epsilon)+\Psi_1(2-\epsilon) + \frac{
\Psi_0(2-\epsilon)-\Psi_0(\epsilon)}{1-\epsilon} \right) .
\label{eq:6.2}
\end{eqnarray}
Note that the right sides of (\ref{eq:6.1}) and (\ref{eq:6.2})
agree, and they are also equal to the right side of
(\ref{eq:5.10}) without the factor $\pi^2\hbar^2/2mL^2$.
Therefore, if we normalise the wave function $\psi_\chi(x,0)$ and
calculate the expectation value of the Hamiltonian, then we find
that for {\it all} values of $\epsilon\in[0,1]$,
\begin{eqnarray}
E_\chi = \frac{\pi^2\hbar^2}{2mL^2}. \label{eq:6.3}
\end{eqnarray}
This is the initial energy of the system. Therefore, we notice the
phenomenon of nonlocal energy transfer and find that the energy is
conserved.

On the other hand, if the position measurement is performed at
time $t>0$, then, as we show in \S\ref{s6}, the infinite energy
density concentrated initially at $x=L-\delta$ will propagate
across the two subchambers. Hence, the nonlocal feature of quantum
mechanics allows the transfer of an infinite amount of energy
across arbitrary large distances provided that we have an infinite
energy source enabling us to insert the barrier instantaneously.

\section{Recognition of the barrier by fractal wave functions}
\label{s6}

After the barrier is in place at $x=L-\delta$, its impenetrability
causes the wave function to vanish at $L-\delta$. It is natural to
ask how the particle becomes aware of this new boundary condition.
We therefore investigate the dynamics of the system in the closed
interval $[0,L-\delta]$. The time-dependent wave function is
\begin{eqnarray}
\psi_\chi(x,t) = \sum_{n=1}^\infty \re^{-\ri E_n t/\hbar} a_n
\chi_n(x) \label{eq:7.1}
\end{eqnarray}
for $0\leq x\leq L-\delta$, where $\chi_n(x)$ and $a_n$ are
specified in (\ref{eq:5.2}) and (\ref{eq:5.4}). The energies are
$E_n = \pi^2\hbar^2 n^2/2m(L-\delta)^2$.

\begin{figure}
{\centerline{\psfig{file=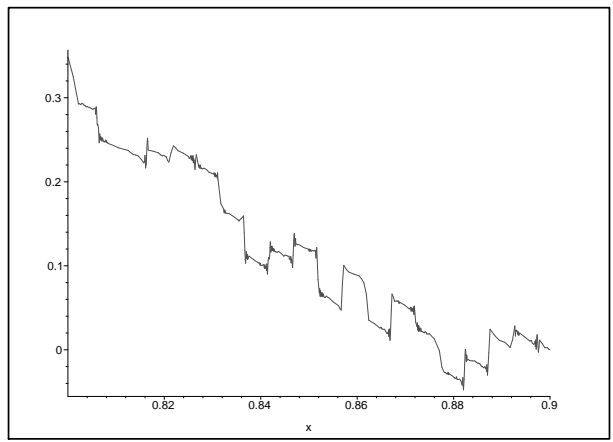,width=8cm,angle=270}}}
\caption{\label{F4} Imaginary part of the wave function for a
particle in the box. This plot is constructed from the first
$50,000$ terms in the Fourier expansion. We have set $L=1$\AA, and
chosen the mass $m_{\rm e}$ of the particle to be the mass of an
electron. The plot shows the wave function in the interval
{0.8\AA~$\sim$~0.9\AA} at time $t=10^{-17}$ seconds. There are
many sharp but {\it continuous} fluctuations in the wave function.
These fluctuations have a self-similar structure and they appear
at all scales.}
\end{figure}

At $t=0$ we have $\lim_{x\to(L-\delta)^-}\psi_\chi(x,0)\neq0$.
This limiting value is nonzero because at $t=0$ there is a
discontinuity at the boundary $x_0=L-\delta$. However, for any
time $t>0$ the dynamics of unitary time evolution respects the new
boundary condition associated with the impenetrable barrier and
forces the wave function $\psi(x,t)$ to vanish at $x=L-\delta$. In
Fig.~\ref{F3} we plot the real and imaginary parts of $\psi(x,t)$
approximated by a truncated Fourier series in the vicinity of the
inserted barrier. This plot shows that the wave function
$\psi(x_0,t)$ does indeed vanish.

Although the discontinuity of the wave function vanishes for
$t>0$, an infinite amount of energy continues to be stored in the
potential well. This energy manifests itself in the form of a
fractal wave function whose energy density in the well is
everywhere infinite. The self-similar structure of the fractal
wave function can be verified by plotting it on successively
smaller intervals while at the same time increasing the number of
terms in the Fourier series expansion. Fig.~\ref{F4} shows the
imaginary part of the wave function $\psi_\chi(x,t)$ at
$t=10^{-17}$ seconds using a $50,000$-term Fourier series. The
energy of the fractal wave function in Fig.~\ref{F4} is infinite
because the probability $p_n$ of finding the particle in the $n$th
energy state decays like $p_n\sim n^{-2}$, while the $n$th energy
level grows like $E_n\sim n^2$.

The fractal structure encountered here has been investigated
previously by Berry (1996). Berry discussed the question of how a
wave function in an infinite potential well evolves in time from a
constant initial state $\psi(x,0)=1/\sqrt{L}$. Berry found that if
an initial wave function in a potential well has a finite jump
discontinuity so that it has infinite energy, then under unitary
time evolution the wave function will instantly develop fractal
structure. Other studies of fractal wave functions have been done
by W\'{o}jcik {\it et al}. (2000), Berry {\it et al}. (2001) and
references cited therein.

\begin{figure}
{\centerline{\psfig{file=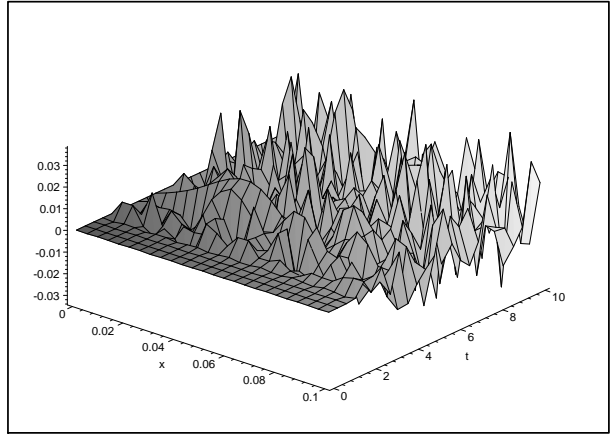,width=8cm,angle=270}}}
\caption{\label{F5} Evolution of the imaginary part of the wave
function. We set $L=10^{10}${\AA} and insert the barrier at $x_0=
0.9\times10^{10}${\AA} rapidly enough to excite the first $1,000$
energy levels. The plot displays the intervals $0\leq x\leq
0.1\times10^{10}$ in units of {\AA} and $10^{-3}\leq t \leq 10$ in
units of seconds. The amplitude of the wave function in the
vicinity of $x=0$ is exponentially small (of order $10^{-4}$) when
$t\sim10^{-3}$ seconds. Therefore, a force on the wall will not be
detected at $x=0$ when $t\ll1$. }
\end{figure}

Although a fractal state such as (\ref{eq:7.1}) requires infinite
energy to create, if we insert a barrier rapidly but in finite
time and with finite energy, then the resulting wave function
displays a fractal-like feature. However, because there is only
finite energy, if a barrier is inserted in the vicinity of one end
of the well, then for large $L$ it will take a while before the
wave function at the other end of the well is excited. An example
in which a barrier is inserted near one end of the well rapidly
enough to excite the first one thousand energy levels is shown in
Fig.~\ref{F5}, where we plot the imaginary part of the wave
function in the vicinity of the other end of the well. As we see
in Fig.~\ref{F5}, there is a time delay before the rise in the
amplitude of the wave function near the well becomes noticeable.

\section{Adiabatic insertion of an impenetrable barrier}
\label{s7}

Inserting an impenetrable barrier into a wave function at a
nodeless point is similar to pushing an object through a viscous
fluid. The work done in the process depends on the speed of
insertion. In \S\ref{s6} we showed that it requires an infinite
amount of work to insert the barrier instantaneously. However, a
slow insertion of the barrier requires only a finite amount of
work.

In this section we consider the case of an {\it adiabatic}
(infinitely slow) insertion of a barrier at a nonnodal point
$x_0$. Inserting a barrier creates a new node at $x_0$ and changes
the state of the system. However, an adiabatic process is one in
which the system does not depart from equilibrium. Thus, during an
adiabatic insertion the change in the energy of the system is
minimised.

\begin{figure}
{\centerline{\psfig{file=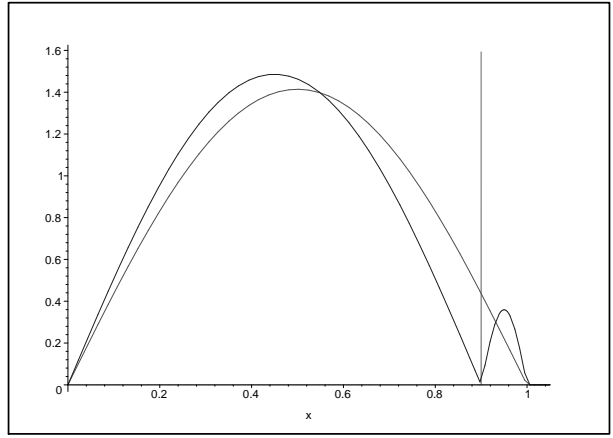,width=8cm,angle=270}}}
\caption{\label{F6} Adiabatic insertion of a barrier at
$x_0=$0.9{\AA} with $L=1$\AA. The initial state $\phi_1(x)$ having
support on $[0,1]$ becomes a pair of (unnormalised) ground states
one on each side of the barrier.}
\end{figure}

We reconsider the configuration in \S\ref{s4}, where a barrier was
inserted into the initial state $\phi_1(x)$ at $x_0=L-\delta$.
However, instead of an instantaneous insertion we now perform an
adiabatic insertion. The states before and after the adiabatic
insertion of the barrier are illustrated in Fig.~\ref{F6}, where
we have taken $\delta=0.1\,L$. The final energy of the system is
the sum of the ground-state energies on both sides of the inserted
barrier, multiplied by the corresponding probability amplitudes.
The energy required to insert the barrier is determined by the
difference between the final and initial energies of the system:
\begin{eqnarray}
\Delta E &=& \frac{\pi^2\hbar^2}{2m} \left( \frac{1}{(L-\delta)^2}
\int_0^{L-\delta} \phi_1^2(x)\rd x + \frac{1}{\delta^2}
\int_{L-\delta}^1 \phi_1^2(x)\rd x - \frac{1}{L^2}\right)
\nonumber \\ &=&  \frac{\pi^2\hbar^2}{2mL^2} \left( \frac{1}
{\epsilon} + \frac{1}{1-\epsilon} + \frac{2\epsilon-1} {\pi
\epsilon^2 (1-\epsilon)^2} \sin(\pi\epsilon) \cos(\pi\epsilon) -1
\right), \label{eq:8.1}
\end{eqnarray}
where $\epsilon=\delta/L$.

For small $\epsilon$ the Taylor expansion of $\Delta E$ is
\begin{eqnarray}
\Delta E\sim\frac{\pi^2\hbar^2}{2mL^2}\Big(2\epsilon
+\twothirds\pi^2\epsilon+3\epsilon^2+{\cal O}
(\epsilon^3)\Big)\quad (\epsilon\ll 1). \label{eq:8.2}
\end{eqnarray}
The term $\frac{2}{3}\pi^2\epsilon$ is a consequence of the cubic
behaviour of the probability in (\ref{eq:5.1}) and is a feature of
low-energy quantum states. High quantum number states do not have
this term and thus this behaviour has no classical analogue.

What happens if we measure the presence or absence of the particle
on one side of the inserted barrier, say, in $[L-\delta,L]$? If
the particle is absent, then the energy that was contained in
$[L-\delta,L]$ will be transferred to $[0,L-\delta]$, where the
particle is now present. However, counter to the intuition one
gains from Fig.~\ref{F6}, the {\it L\"uders state} (the state
having support on $[0,L-\delta]$ that results after the wave
function collapses) is not a normalised ground state.

To show that the L\"uders state is not the ground state we assume
the contrary. If it were the ground state, then the final energy
of the system would be proportional to $(L-\delta)^{-2}$. The
initial energy of the system before the adiabatic insertion of the
barrier was proportional to $L^{-2}$. The energy difference of
these two ground states must equal the energy $\Delta E$ required
to insert the barrier. However, the energy $\Delta E$ required to
insert the barrier is greater than this difference except when the
insertion is located at the centre $\delta=L/2$:
\begin{eqnarray}
\frac{1}{(L-\delta)^2} \int_0^{L-\delta} \phi_1^2(x)\rd x +
\frac{1}{\delta^2} \int_{L-\delta}^1 \phi_1^2(x)\rd x
-\frac{1}{L^2} \geq \frac{1}{(L-\delta)^2} -\frac{1}{L^2},
\label{eq:8.3}
\end{eqnarray}
where equality holds only if $\delta=L/2$. Thus, the energy
injected into the system by the adiabatic insertion of the barrier
at $\delta\neq L/2$ is so large that after the absence of the
particle on one side of the barrier is confirmed, the wave
function on the other side will be excited to a state higher than
the ground state.

For $\epsilon\ll 1$ the right side of (\ref{eq:8.3}) becomes
\begin{eqnarray}
\frac{1}{(L-\delta)^2} - \frac{1}{L^2} \sim \frac{1}{L^2} \Big(
2\epsilon + 3\epsilon^2 + {\cal O}(\epsilon^3)\Big).
\label{eq:8.4}
\end{eqnarray}
This result shows that the term $\frac{2}{3}\pi^2\epsilon$ in
(\ref{eq:8.2}) is the additional energy that excites the system
above the ground state when the absence of the particle in the
interval $[L-\delta,L]$ is confirmed. This additional term arises
whenever we insert the barrier at a location away from the centre
of the potential well that contains a low-energy initial wave
function.

It is interesting to note that depending on whether a measurement
to confirm the presence or absence of the particle in the region
$[L-\delta, L]$ is conducted, the expectation value of the
Hamiltonian in the region $[0,L-\delta]$ takes different values
when $\epsilon\ll1$. Specifically, if the system has not been
disturbed by a measurement, then to first order in $\epsilon$ we
have
\begin{eqnarray}
\int_0^{L-\delta} \phi_1(x) H \phi_1(x) \rd x =
\frac{\pi^2\hbar^2} {2mL^2} \big( 1+2\epsilon \big) .
\label{eq:9.1}
\end{eqnarray}
On the other hand, if the system has been disturbed by a
measurement, then the energy expectation becomes
\begin{eqnarray}
{\rm tr}(\rho H) = \frac{\pi^2\hbar^2} {2mL^2} \Big( 1 + \left( 2+
\twothirds\pi^2\right) \epsilon \Big). \label{eq:9.2}
\end{eqnarray}
Here the density matrix $\rho$ reflects the two possible outcomes
of the measurements, whose statistics is provided by
(\ref{eq:5.1}). For example, for $\epsilon$ of order $1\%$ there
is about a $6\%$ difference between the energy expectation values.

\section{Maxwell's daemon for square wells}
\label{s8}

A quantum version of Maxwell's daemon was considered by numerous
researchers (Zurek 1986, Bennett 1987, Lloyd 1997, and references
therein). In the context of the square-well systems considered in
this paper, the Maxwell daemon scenario is as follows: A barrier
is inserted adiabatically at the centre of a well containing the
initial ground-state wave function $\phi_1(x)$. A node appears at
the centre, and on each side of the node the wave function is in
the (unnormalised) ground state. The daemon then performs a
measurement to determine on which side of the barrier the particle
is present and in doing so causes the wave function to collapse.
Until now, the barrier considered in this paper was fixed in
place; however, in this scenario the barrier is a classical object
that is allowed to slide to the left or right inside the well. The
daemon then releases the barrier, and as the barrier slides and
the subchamber containing the particle expands, energy is
extracted from the collapsed wave function. This energy can be
used to reduce the entropy of the environment. When the barrier
reaches the edge of the well, the initial wave function is
recovered. This appears to be a cyclic engine that violates the
second law of thermodynamics.

The standard argument (see Landauer 1961 and Zurek 1986) to show
why the second law is actually not violated is that to close the
cycle of this engine completely the daemon must erase the
information about where the particle was when the barrier was
inserted. Because the probability of finding the particle in a
given subchamber was $\frac{1}{2}$, the amount of this information
is $\ln2$. The act of erasing the memory balances the decrease in
the entropy (the so-called Landauer's principle) so that there is
no violation of the second law.

If the barrier is not inserted at the centre of the well, the
entropy associated with erasing the memory is reduced. This is
because the probability of finding the particle in a given
subchamber is not $\frac{1}{2}$. Classically, this reduced entropy
still balances the reduction of the entropy in the environment.
Similarly, for a quantum system in equilibrium with a high
temperature heat reservoir, as considered by Zurek, Landauer's
erasure principle also applies without modification because no
energy is needed to insert the barrier.

However, for small $\epsilon$ and for low-energy quantum states
the entropy reduced in the environment no longer balances the
information contained in the daemon's memory. Specifically, the
probabilities of trapping the particle in the subchambers are
$p(\epsilon)$ and $1-p(\epsilon)$, where $p(\epsilon)$ is given in
(\ref{eq:5.1}). The entropy lost by the daemon can be expressed as
a series expansion for $\epsilon\ll1$:
\begin{eqnarray}
S(\epsilon)\sim\left[\twothirds\pi^2\Big(1-
\ln\left(\twothirds\pi^2\right)\Big) -2\pi^2\ln(\epsilon) \right]
\epsilon^3,
\end{eqnarray}
which approaches zero rapidly as $\epsilon\to0$. On the other
hand, the energy that the daemon has extracted is of order
$\epsilon$.

This apparent violation of the second law is avoided by
recognising that inserting the barrier adiabatically at a nonnodal
point requires a finite amount of work. Therefore, the usual
explanation, which relies solely on Landauer's principle, for why
the second law is not violated is insufficient in such
circumstances. We have shown here that Landauer's principle when
used in this quantum context must be augmented by including the
energy required to insert the barrier.

\section{Fractal states with unbounded entropy}
\label{s9}

Let us examine the entropy of fractal quantum states. The term
{\it entropy} used here is the {\it von Neumann entropy}
associated with the density matrix that represents the mixed state
of the system after a measurement (of the energy, for example) is
performed on the initial pure state. A pure state has zero
entropy, so when we speak of the entropy of a fractal wave
function, it is always understood to indicate the von Neumann
entropy associated with the density matrix after energy
measurement.

The fractal wave functions encountered here and elsewhere in the
literature share the property that the associated entropy is
finite. It is possible, however, to construct a special class of
fractal wave functions whose entropies are infinite. To show that
the entropy associated with states generated by an instantaneous
insertion of the barrier is finite, we recall that the probability
$p_n$ of finding the system in the $n$th energy level is given for
large $n$ by $p_n\sim n^{-2}$. This follows directly from
(\ref{eq:5.4}), and the same $n$-dependence applies to Berry's
fractal states (1996). Therefore, the von Neumann entropy
$S=-\sum_n p_n\ln p_n$ is finite. In the case of fractal states
considered by W\'ojcik {\it et al}. (2000), the probability $p_n$
decays exponentially fast, and thus the associated von Neumann
entropy converges rapidly.

\begin{figure}
{\centerline{\psfig{file=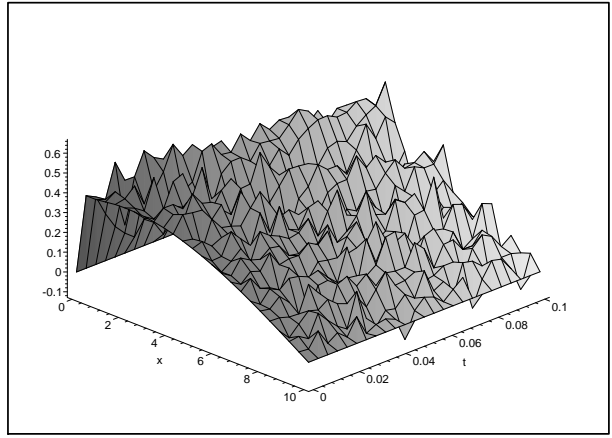,width=8cm,angle=270}}}
\caption{\label{F7} Evolution of the real part of the wave
function (\ref{eq:10.2}), approximated by the first {10,000} terms
in its Fourier expansion. We set $\nu=0.5$, $L=10$\AA, and the
mass of the particle is set to the mass $m_{\rm e}$ of an
electron. The plot is given for the intervals $0\leq x \leq 10$ in
units of {\AA} and $0<t\leq 10^{-16}$ in units of seconds.}
\end{figure}

As noted above, we can create a new class of fractal wave
functions for which the associated entropy is infinite. Consider
as an example the probability
\begin{eqnarray}
p_n = \frac{1/N(\nu)}{(n+\nu)[\ln(n+\nu)]^2}, \label{eq:10.1}
\end{eqnarray}
where $\nu>0$ is a constant and $N(\nu)$ is the normalisation. The
probability (\ref{eq:10.1}) is an example of a fat-tail
distribution for which none of the associated moments exist and,
similarly, the associated entropy is also infinite. The key idea
here is that the norm of the state is finite because the integral
$\int^\Lambda 1/x(\ln x)^2\rd x=1/\ln\Lambda$ is finite as
$\Lambda\to\infty$, while the entropy is infinite because the
integral $\int^\Lambda 1/x\ln x\,\rd x=\ln(\ln\Lambda)$ diverges
as $\Lambda\to\infty$.

Letting $c_n=\sqrt{p_n}$, we can construct a wave function
representing a state of a particle in the potential well:
\begin{eqnarray}
\psi(x,t)=\sum_{n=1}^\infty \re^{-\ri E_n t/\hbar}c_n \phi_n(x).
\label{eq:10.2}
\end{eqnarray}
At $t=0$ this wave function is smooth for $x>0$ and singular at
$x=0$. Such an initial state might be created by an instantaneous
compression of the potential well at $x=0$. The key difference
between the wave function (\ref{eq:10.2}) and the states
considered in \S\ref{s6} is that at the boundary $x=0$ the initial
wave function $\psi(x,0)$ (\ref{eq:10.2}) is infinite. However,
once the system evolves in time, the wave function satisfies the
homogeneous boundary condition at $x=0$ and becomes fractal. An
example is displayed in Fig.~\ref{F7}.

\begin{figure}
{\centerline{\psfig{file=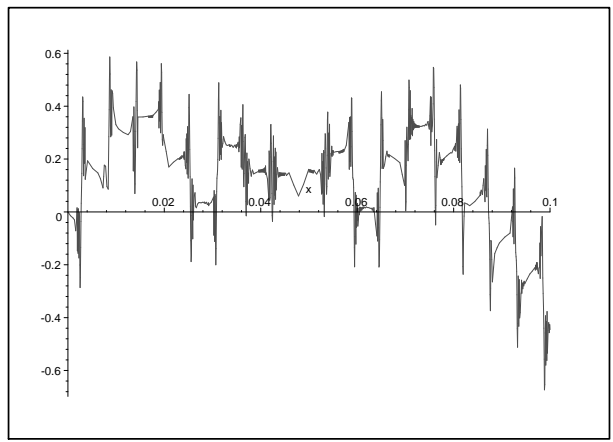,width=8cm,angle=270}}}
\caption{\label{F8} Ringing phenomenon of the wave function
(\ref{eq:10.2}), approximated by the first {20,000} terms in its
Fourier expansion. Whenever time reaches $t=2km_{\rm e}L^2/
\pi^2\hbar$ $(k=1,2,3,\ldots)$, the wave function takes a
characteristic form as shown here. We set $\nu=0.5$, $L=1$\AA, and
the plot is given for the interval $0\leq x \leq 0.1$ in units of
{\AA}, and time is set so that $\pi^2\hbar t/2m_{\rm e} L^2=1$
(about $10^{-15}$ seconds). }
\end{figure}

One particularly interesting characteristic behaviour associated
with the wave function $\psi(x,t)$ of (\ref{eq:10.2}) is that
whenever the time variable $t$ reaches the values
\begin{eqnarray}
t=\frac{2kmL^2}{\pi^2\hbar}\quad (k=1,2,3,\ldots), \label{eq:10.3}
\end{eqnarray}
the wave function exhibits a peculiar ringing phenomenon as
illustrated in Fig.~\ref{F8}. There are 117 oscillations of the
kind shown in the figure in the interval $[0,L]$.

If we measure the energy of the system prescribed by the wave
function (\ref{eq:10.2}), the entropy of the resulting mixed state
is infinite even though the size $L$ of the system is finite. This
seems to imply that quantum mechanics is not consistent with black
hole physics, for which the entropy of a system is bounded by the
surface area of a black hole (see Bekenstein 1994 and references
cited therein). However, recall that the energy of the present
system is also infinite. Thus, in order to create an infinite
entropy state for a particle trapped in a potential well, the
system must also possess infinite energy. One may argue that when
the energy of the system reaches a threshold value, the system
collapses to form a black hole. As a consequence, the entropy of
the system cannot diverge. On the other hand, there are other
systems, such as the bound states of a hydrogen atom, for which
energy expectation is always finite. In \S\ref{s10} we examine
some other quantum systems having unbounded entropy.

\section{Limits on quantum states having infinite entropy}
\label{s10}

One can construct wave functions using the probability weights
(\ref{eq:10.1}) for other quantum systems, such as a harmonic
oscillator or a hydrogen atom. What can we say about the energy
and size of such systems?

We investigate first the properties of a particle trapped in a
harmonic potential $V(x)=\frac{1}{2}m\omega^2x^2$. The eigenstates
of the Hamiltonian (with $m\omega/\hbar=1$) are
\begin{eqnarray}
\phi_n(x) = \frac{\pi^{1/4}}{\sqrt{2^n n!}} \re^{-\frac{1}{2}x^2}
H_n(x), \label{eq:11.1}
\end{eqnarray}
where $H_n(x)$ denotes the $n$th Hermite polynomial. Using these
eigenstates we construct an initial wave function $\psi(x)=\sum_n
c_n \phi_n(x)$.

\begin{figure}
{\centerline{\psfig{file=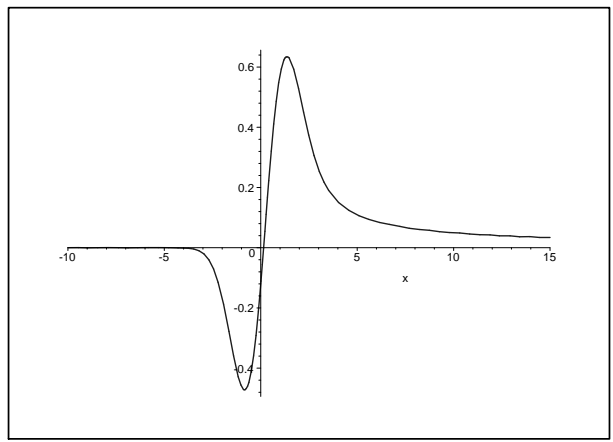,width=8cm,angle=270}}}
\caption{\label{F9} Initial wave function $\psi(x)$ constructed
from the weights (\ref{eq:10.1}) and the eigenstates
(\ref{eq:11.1}). The wave function is smooth and square
integrable. However, its square $\psi^2(x)$ possesses a fat-tail
distribution. Asymptotically, we have $\psi(x)\sim (\sqrt{x} \ln
x)^{-1}$ as $x\to+\infty$.}
\end{figure}

Unlike the case of a particle in a square well, the wave function
$\psi(x)$ exhibits no discontinuity or fractal structure, as is
evident from the plot of the wave function in Fig.~\ref{F9}. This
is because the initial wave function $\psi(x)$ does not violate
the boundary conditions associated with the Schr\"odinger
eigenvalue problem. Nonetheless, if an energy measurement is
performed on the system prepared in the state $\psi(x)$, the
resulting density matrix has infinite von Neumann entropy.
Similarly, the energy of the system is also infinite because the
expectation value of the Hamiltonian diverges. The energy
expectation value can be determined either by taking the sum
$\sum_n p_n E_n$, where $E_n\sim n$ ($n\to\infty$), or by
considering the asymptotic behaviour of the wave function
$\psi(x)$. Specifically, numerical studies indicate that
\begin{eqnarray}
\psi(x)\sim\frac{0.44}{\sqrt{x}\ln x} \label{eq:11.2}
\end{eqnarray}
as $x\to\infty$. Therefore, the probability distribution
$\psi^2(x)$ associated with the position of the particle has a fat
tail and none of the moments exists. In particular, the
expectation value of $x$ is infinite, so the characteristic size
of the system is infinite.

Let us now consider the case of a Coulomb potential for which the
energy eigenvalues $E_n$ are proportional to $-1/n$. In this case,
the expectation value of the Hamiltonian is necessarily bounded
for an arbitrary state. Therefore, it is possible to create a
mixed state using bound states of the hydrogen atom for which the
entropy is infinite, while keeping the system energy finite.
However, the expectation value of $r$ in the $n$th Coulombic bound
state grows like $n^2$ for large $n$ (Messiah 1958). Therefore,
the expectation value of the radius $r$ in the infinite-entropy
state is infinite. Hence, the surface area of the mixed state with
diagonal elements $p_n$ given by (\ref{eq:10.1}), is infinite.

With these observations we conjecture that {\it for a quantum
state to have infinite entropy, either its energy or size, or
possibly both, must also diverge}. If this indeed is the case,
then we may argue that quantum theory is not in direct
contradiction with laws of black hole thermodynamics.

\section{Effects of separating the subchambers}
\label{s11}

Can the nonlocality property of quantum mechanics described in
this paper be used to communicate rapidly over great distances?
Recall that diffusion is a spreading process that proceeds
infinitely fast. The Schr\"odinger equation is a diffusion
equation, so an initial wave function having compact support will
evolve immediately into a wave function that is nonvanishing over
all space. Nevertheless, at great distances the wave function is
exponentially small, and it is impossible to take advantage of the
nonlocality of the diffusion process to send signals.

However, if the support of the wave function remains finite, one
may ask if instantaneous communication now becomes practical: Can
we use the configuration of two widely separated square wells and
the entanglement of states to transmit signals instantaneously?

Suppose we prepare many identical copies of a system of a single
particle trapped in a square well and insert impenetrable barriers
into each of the potential wells. One subregion is taken to
observer $A$ and the other taken to observer $B$. This establishes
a `telephone line' (cf. Gisin 1990) between $A$ and $B$. Suppose
that at an agreed time $A$ wishes to transmit one bit information
to $B$. To do so $A$ will either measure the presence of the
particles or $A$ will do nothing. If observer $B$ can determine
statistically whether $A$ has performed the measurement, then one
bit of information is transmitted.

Considerations of such a problem require careful analysis of the
effects of the physical displacement of the subchambers on the
wave functions contained therein. If the subchambers are separated
adiabatically, then by definition this separation will not affect
the wave functions. However, such a separation requires an
infinite amount of time. As a consequence, even if an
instantaneous communication across long distances were possible
(for example, using the discrepancy of the energy expectations in
(\ref{eq:9.1}) and (\ref{eq:9.2})), it requires an infinite amount
of time to establish the telephone line.

To overcome this problem we could try to separate the boxes in
finite time. However, in this case we believe that the wave
function collapses to a density matrix because accelerating the
box\footnote{ In general, the equivalence principle implies that
if a box containing a particle is accelerated, there is an induced
gravitational force on the particle. Thus, during acceleration and
deceleration the floor of the square well is not horizontal;
rather, it slopes linearly downward to the left or right, and the
wave function in the well is an Airy function.} constitutes a
measurement of whether there is a particle in the box. If it were
possible to separate the boxes in a finite time without collapsing
the wave function, then it would be possible to send signals
instantaneously over great distances. Moreover, since the
accelerated box can extract energy from its environment, it is
possible to transmit an arbitrarily large quantity of energy
instantaneously across large distances.

\section{Discussion}
\label{s12}

The system considered here is perhaps the simplest quantum system
represented in terms of an infinite-dimensional Hilbert space.
Nevertheless, by a thorough analysis of the various effects of
dividing the system into a pair of subsystems we were led to a
number of interesting and peculiar features of quantum mechanics.
Some of the gedanken experiments considered in this paper might
well be feasible using the procedures recently developed by
Konstantinov and Maris (2003).

Because an insertion of the barrier in a potential well creates
entangled subsystems, one may ask whether it is possible to
formulate a version of the Bell inequality that is appropriate for
this configuration. The Bell inequality relies on the statistics
of noncommuting observables. In the present example one might
consider the measurement of the presence or absence of a particle
in a subchamber and the energy contained in a subchamber as the
relevant observables.

In the discussion of a quantum-mechanical Maxwell daemon we found
an example in which the standard argument based on Landauer's
principle is insufficient. Nonetheless, we were able to resolve
the problem by taking into account the energy needed to insert the
barrier into the potential well.

We have found a new class of fractal states having infinite
entropy. Our discussion in \S\ref{s10} suggests that if a quantum
state has infinite entropy, then it must also have infinite energy
and/or infinite size. This raises the interesting question of
whether it is possible to derive a bound on entropy or perhaps
even a bound on the gravitational constant, conditional on finite
size and energy, in the context of nonrelativistic quantum
mechanics.

The discussion in \S\ref{s11} concerning the mechanical
displacement and separation of the subchambers leads to an
interesting gedanken experiment to test the so-called Wigner
interpretation of quantum mechanics. According to this
interpretation, the collapse of the wave function occurs only when
the result of a measurement has been registered by a conscious
being (Wigner 1963). Therefore, if the subchambers are displaced
by a mechanical device that does not involve conscious awareness,
and as a result the wave function does not collapse, then it is
theoretically possible to transport an unbounded quantity of
energy instantaneously over large distances.

\begin{acknowledgements}
CMB is supported by the U.S.~Department of Energy, the U.K. EPSRC,
and the John Simon Guggenheim Foundation. DCB is supported by The
Royal Society.
\end{acknowledgements}

\end{document}